\documentclass{article}
\usepackage[utf8]{inputenc}
\usepackage{graphicx}
\usepackage{bbm}
\usepackage{lineno}
\usepackage{csquotes}
\usepackage[breaklinks=true]{hyperref}
\usepackage[english]{babel}
\usepackage{multirow}
\usepackage[outdir=./]{epstopdf}
\usepackage[labelfont=bf]{caption}
\usepackage{mathrsfs}
\usepackage{adjustbox}
\usepackage{algorithm}
\usepackage{siunitx}
\usepackage{tabularx}
\usepackage{nag}
\usepackage{hvfloat}
\usepackage{booktabs}
\usepackage{lipsum}
\usepackage{authblk}
\usepackage{url}

\usepackage{amsmath,amsfonts}
\usepackage{multicol}
\usepackage{multirow}
\usepackage{amssymb}
\usepackage{mathtools}
\usepackage{colortbl}
\usepackage{rotating}
\usepackage{epsfig}
\usepackage{breakcites}
\usepackage{color,soul}
\usepackage{xcolor}
\usepackage{xurl}

\title{Characterization of flexible electricity in power and energy markets}
\author[1]{G\"uray Kara\footnote{Corresponding author-Email:guray.kara@ntnu.com}}
\author[1]{Asgeir Tomasgard}
\author[2]{Hossein Farahmand}
\date{ }	
\affil[1]{{\footnotesize Norwegian University of Science and Technology, Department of Industrial Economics and Technology Management}}		
\affil[2]{{\footnotesize Norwegian University of Science and Technology, Department of Electric Power Engineering}}

\begin{document}
\maketitle		
%	\begin{frontmatter}	
\begin{abstract}
	The authors provide a comprehensive overview of flexibility characterization along the dimensions of time, spatiality, resource, and risk in power systems. These dimensions are discussed in relation to flexibility assets, products, and services, as well as new and existing flexibility market designs. The authors argue that flexibility should be evaluated based on the dimensions under discussion. Flexibility products and services can increase the efficiency of power systems and markets if flexibility assets and related services are taken into consideration and used along the time, geography, technology, and risk dimensions. Although it is possible to evaluate flexibility in existing market designs, a local flexibility market may be needed to exploit the value of the flexibility, depending on the dimensions of the flexibility products and services. To locate flexibility in power grids and prevent incorrect valuations, the authors also discuss TSO-DSO coordination along the four dimensions, and they present interrelations between flexibility dimensions, products, services, and related market designs for productive usage of flexible electricity.
\end{abstract}
	
%\end{frontmatter}
			
\section{Introduction}	
\label{sec:intro}
The increasing share of variable renewable energy sources (VRES) introduces short-term uncertainty and variability in power systems. This creates a need for flexibility in order to maintain a continuous supply-demand balance. VRES affect electricity prices, power quality, and uncertainty in power-generation processes. Flexibility derived from power and energy resources can be useful for balancing the effects of VRES on power markets. To exploit the flexibility, power markets should provide incentives for optimal valuation of flexibility for both short-term purposes (operations) and long-term purposes (investments). In this paper we study the characterization of flexibility in energy systems along four dimensions: time, spatiality (space), resource, and risk. The products and services of the flexibility are examined in existing market designs and for possible new market designs.

The main contribution of this paper is an overview of dimensions for flexibility characterization in order to give insights into its usage in different market designs and on different time horizons. There is not a unified definition of flexibility, but in this study we focus on the following definition: \textquotedblleft\textit{Flexibility is the modification in the generation and/or consumption pattern of electricity according to an external signal in order to meet energy system needs}"~(\cite{mandatova2014}, p.5). Based on this definition, the flexibility can be demand-side flexibility and supply-side flexibility. In addition, we observe flexibility provision from power grid and storage resources. Sæle et al. (2018) state that the utilization of flexible resources by system operators is not just prioritized due to changes in demand and generation~\cite{saele2018utilization}. Therefore, we discuss the concept of flexibility (e.g.,~\cite{mit2016utility}), by characterizing it in terms of four main dimensions: time, spatiality, resource technology, and risk profile.  

The spatial characteristics of a flexibility resource are important, as some flexible resource technologies are relevant only on a local scale, whereas others are usable system-wide. The location of the flexibility resource can affect flexibility trading and the flexibility effectiveness of the services provided by transmission system operators (TSOs) and distribution system operators (DSOs)~\cite{kouzelis2015geographical, us2015}. The time dimension has a strong impact on flexible resources. All flexible resources have a response time, duration, and ramping characteristics. Services and products based on flexibility need to be designed according to the time dimension. The resource technology, such as supply-side flexibility, originates from power plants (centralized suppliers). Demand-side flexibility technology originates from the behaviors of industrial and household users of electricity. According to the recent developments in information and communication technologies (ICT), communication between end users and power generators is leading to demand-side flexibility. Although there is more ongoing collaboration with industrial users for flexibility, such as load curtailment, also participation by households has been motivated~\cite{mandatova2014,winterpackage2016}. Storage-side flexibility technologies are suitable for flexibility provision due to their availability in time and amount with limited capacity and charge/discharge rates. Grid flexibility is one of the key elements in electricity systems because the grid infrastructure for transmission and distribution of electricity allows for increases or decreases in the degree of flexibility. The risk profile of a flexible resource is related to the uncertainty in the flexibility provision process.

Our research provides a starting point for discussions of the concept of flexibility, market designs, and other flexibility characterization schemes. Moreover, we discuss the importance of DSO-TSO interactions for flexibility trading, and we relate this to the four flexibility dimensions. In this paper, we focus on the following:

\begin{enumerate}
	\item The characterization of flexibility in power and energy systems in terms of the spatiality, time, resource, and risk dimensions
	\item The efficiency and suitability of existing and possible new power/energy markets for the usage of the flexibility.
	\item Pricing of the flexibility in different market designs.
	\item Different flexibility-based products and services along the four dimensions in market designs.
	\item Presentation and discussion of TSO-DSO coordination for flexibility-based services.
\end{enumerate}

The paper is structured as follows. Section 2 discusses the dimensions of flexibility as time, spatiality, and technology. Section 3 describes the risk dimension of flexibility and related market designs. Section 4 explains the flexibility products. Section 5 discusses new market designs for flexibility trading and DSO-TSO coordination. Section 6 presents the conclusions.

\section{The dimensions of flexibility}
\label{sec:dime}

Flexibility characterization is significant for future flexibility trading mechanisms and market designs. The flexibility itself is not a homogeneous product; it is context-dependent and relates to the characteristics of various energy resources that should be represented with multiple dimensions (to provide granularity)~\cite{mit2016utility}. Several attempts have been made to characterize flexibility for the purpose of characterization. For example,~\cite{de2015organizing} identify the four dimensions of flexibility as three physical dimensions—infrastructure, geography, interfaces with other energy carriers’ dimensions—as well as the time dimension, namely the relevant time scale. Furthermore, they show how a market design should be adapted to these dimensions by considering the changing roles of participants in the system. However,~\cite{de2015organizing} do not discuss the interrelations between the dimensions.

\cite{eid2016managing} describe four dimensions mainly in relation to their technology. According to them, resources for flexibility are distributed energy resources (DERs), such as electric vehicles (EVs), combined heat and power (CHP) units, and electric water heaters. The four dimensions are the amount of power, the moment of provision, duration, and specific location of resources. Although the study by~\cite{eid2016managing} is quite informative, only DERs are considered as flexible resources. They do not discuss the response time, flexibility resources other than DERs, and risk in flexibility provision. By contrast,~\cite{ela2016wholesale} propose three dimensions of flexibility characterization: absolute power output capacity range (MW), speed of power output change (MW/min), and the duration of energy levels (hours of MW). However, they do not discuss the spatiality dimension.

In this paper, inspired by the \textit{Nordic market balancing concept}~\cite{statnett2017nordic}, we suggest four dimensions for flexibility characterization: resource, spatiality, time, and risk dimensions. According to the concept, in energy and power markets there are three prerequisites for flexibility provision and utilization of the flexibility: the service type, the time of need, and the location of the balancing. In this paper, we extend these prerequisites and relate them to the risk dimension for a holistic understanding of flexibility in different market designs.

\subsection{The spatiality dimension of flexibility}
\label{ssec:spatial}
When looking at spatiality in a realistic sense, it is trivial to observe that the location of any physical product is important for logistics. The price and provision of a product is related to where it is produced and consumed. Hence, in electricity transmission and distribution, especially for flexibility usage, the geography of the flexibility resource is important. The geographical location of a resource is important for the type of flexibility product that can be delivered, including both reactive and active power. Transmission and distribution of active power over long distances is possible, but for reactive power it is inefficient due to high network losses.

\begin{figure}[ht!]
	\centering
	\includegraphics[width=0.7\linewidth]{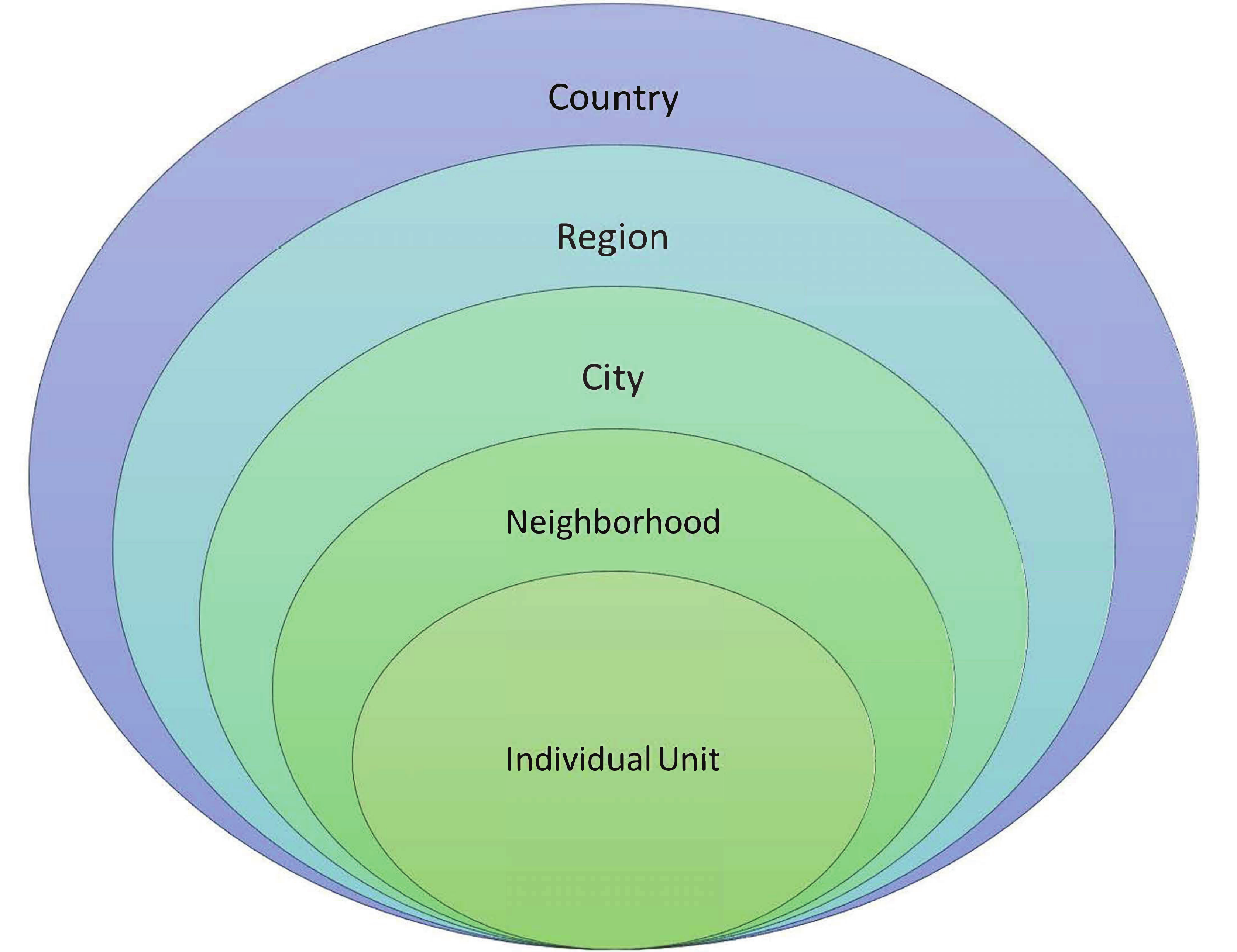}
	\caption{Spatiality dimension.}
	\label{fig:spatial}
\end{figure}

During times of grid congestion, the location of available flexibility will affect the decision-making process. In addition, for a location with a need for TSO-DSO interaction, some resources may be used both by the DSO in the distribution grid and by the TSO for the transmission grid. The possible geographies of flexibility provision are illustrated in Figure~\ref{fig:spatial}.

\subsection{The time dimension of flexibility}
\label{ssec:time}

\begin{figure}[ht!]
	\centering
	\includegraphics[width=0.8\linewidth]{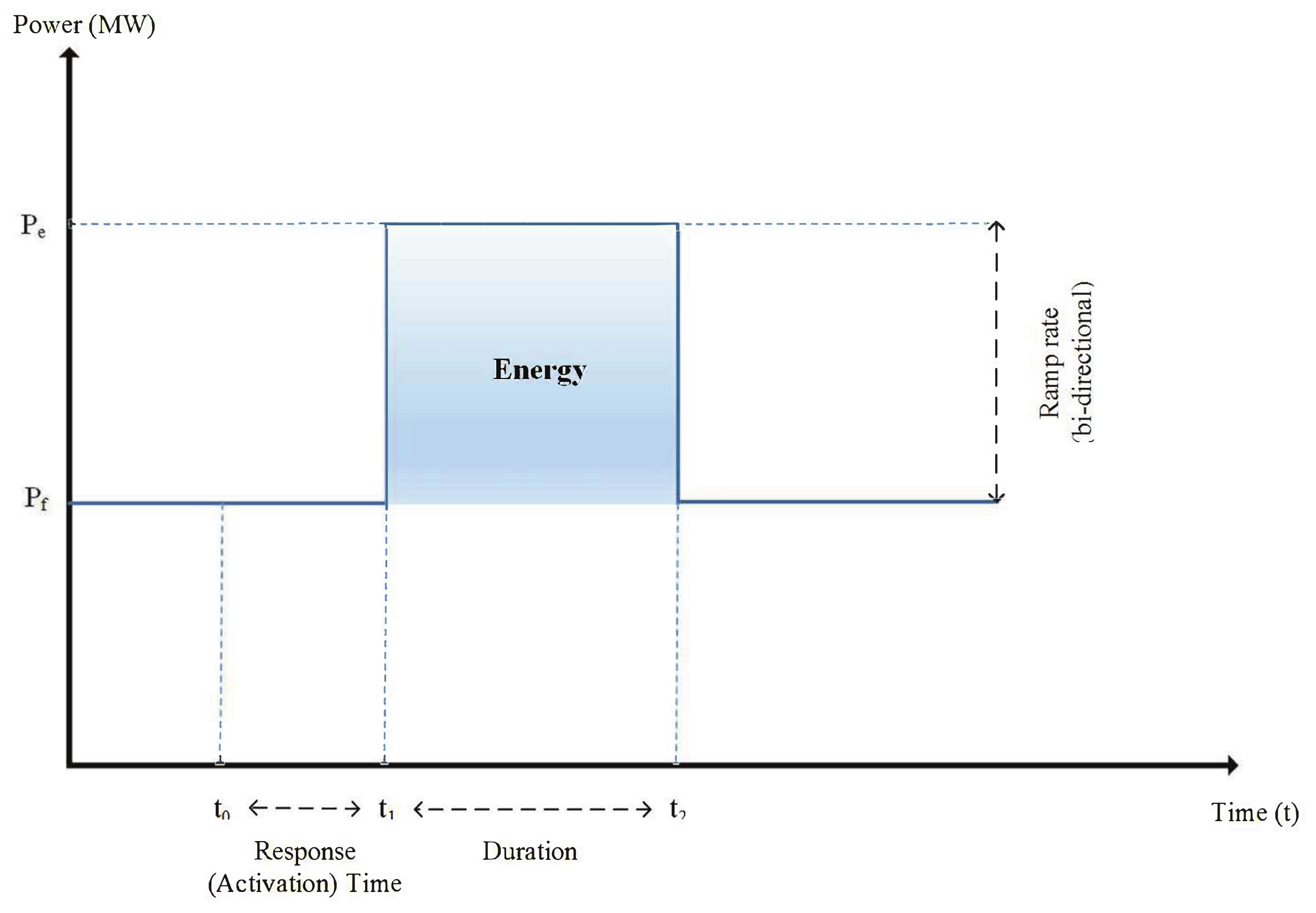}
	\caption{Characteristics of flexibility in system-wide scale~\cite{eid2016managing}.}
	\label{fig:flexchar}
\end{figure}

Based on technological characteristics, market design, and system architecture, the time dimension can be divided into four subdimensions: activation time, ramping rate, duration time, and market time resolution. The activation time concerns how quickly the flexible resource becomes available for usage. The activated flexibility could be useful in a specific time interval (i.e., the duration). The ramping rate of the flexibility resource refers to how fast flexibility resource can ramp-up or ramp-down. Especially in the case of market designs with short time horizons, the ramping rate of a resource should be fast due to the immediate need for power. Based on~\cite{eid2016managing} three subdimensions (activation time, ramping rate, duration time) are illustrated with some modifications in Figure~\ref{fig:flexchar}.

In Figure~\ref{fig:flexchar} the difference between $P_{e}-P_{f}$ is the ramp rate—how fast power can be increased or decreased. The symbol $t$ represents time, and $t_0$, $t_1$, and $t_{2}$ respectively symbolize the signaling, starting, and stopping time of flexibility. The difference between $t_{1}-t_{0}$ is the response (activation) time of the flexibility, while $t_{2}-t_{1}$ is the duration of the flexibility.

\begin{figure}[ht!]
	\centering
	\includegraphics[width=0.8\linewidth]{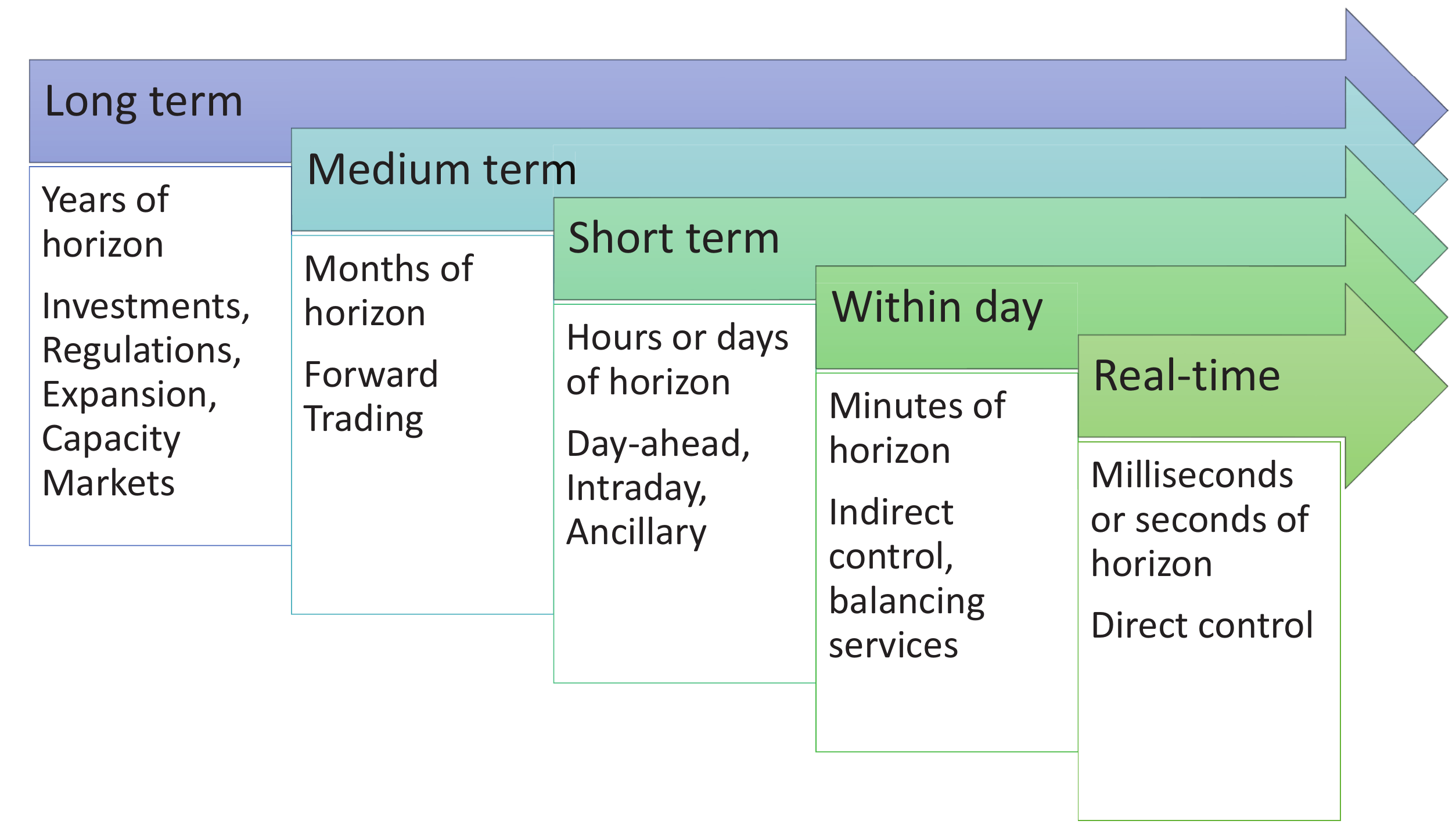}
	\caption{Flexibility trading horizons and markets.}
	\label{fig:time}
\end{figure}

The fourth subdimension of time concerns the relevant market horizon. Different market designs are based upon various time intervals and customer needs~\cite{hillberg2019flexibility}. Hence, the flexibility provision process should be considered with similar time-related decision-making. Different time properties of resources make it possible to participate in different markets for multiple purposes, such as ancillary services for restoring the quality of power in a grid. It is possible to observe different flexibility resources with relevance from milliseconds to years. The structure of the time dimension with respect to flexibility trading horizons and markets is shown in Figure~\ref{fig:time}.
 
The time dimension may be the most important dimension for flexibility and its usage. According to the results of a survey of industry players (managers and modelers) conducted by~\cite{helms2016timing}, with an accurate timing strategy, timing-based flexibility business models in the energy sector could increase their profits while reducing their downside risk. The timing of the market participant could differ for supply-side flexibility resources compared with demand-side flexibility resources. A system operator or a market participant could either use only a single flexibility resource with a single timing strategy or they could harvest multiple resources and have a time coupled portfolio of flexibility.

\subsection{The resource dimension of flexibility}
\label{ssec:res}
\begin{figure}[ht!]
	\centering
	\includegraphics[width=0.8\linewidth]{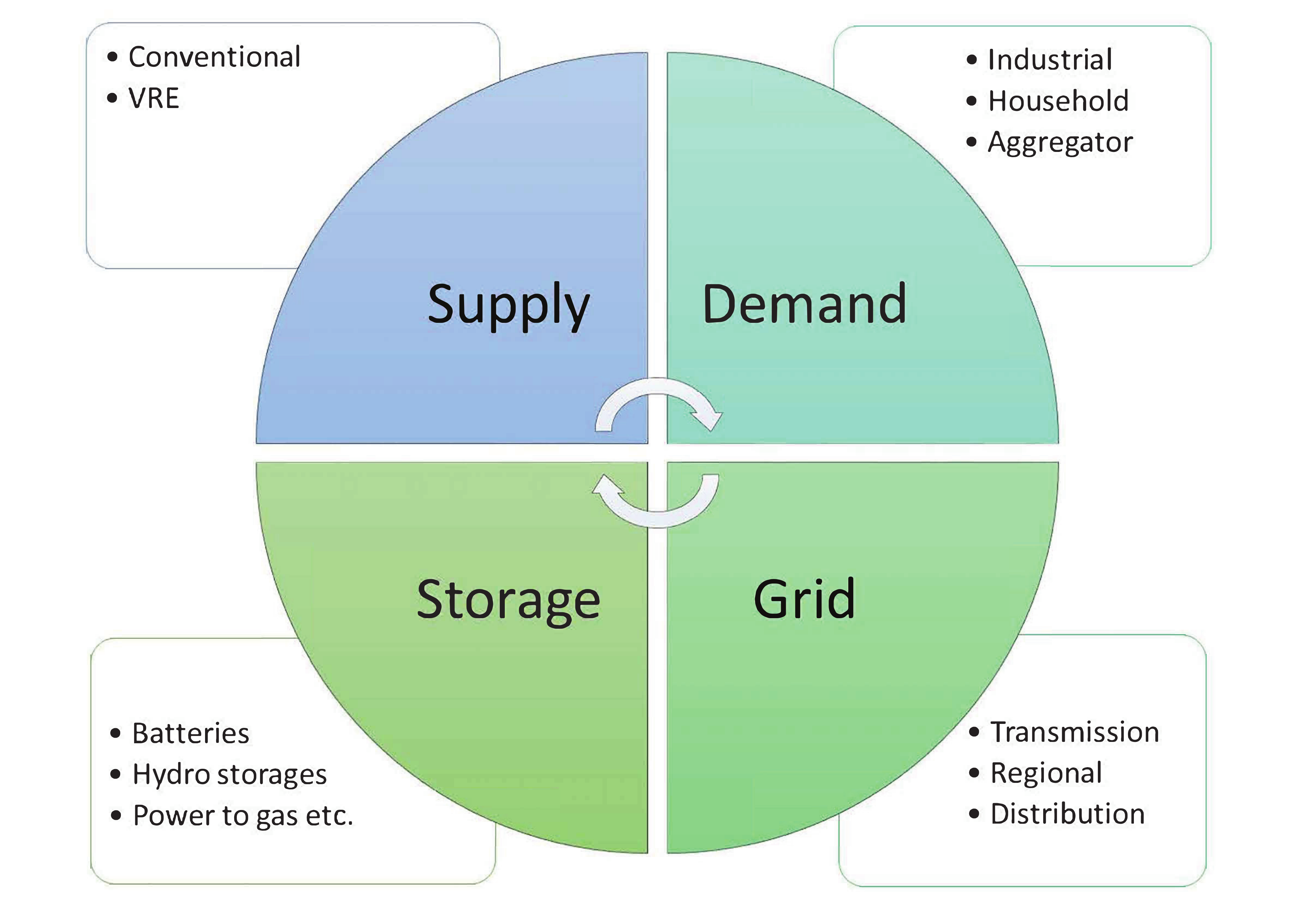}
	\caption{Resource dimension.}
	\label{fig:resource}
\end{figure}

The resource type and the technology of a flexibility asset might vary with different time horizons and locations. Moreover, the marginal cost of generation and the marginal utility of the flexibility provider are related to its resource technology. In this context, we consider four major flexibility resources based upon their technologies: supply side, demand side, grid side, and storage side. The four resource technologies are represented in Figure~\ref{fig:resource}.

\subsubsection{Supply-Side flexibility}
\label{sec:supflex}

The traditional flexibility in power systems originates from either ramp-up or ramp-down of conventional power plants. Variability in the load (demand) profile is the primary reason for conventional usage~\cite{international2011harnessing, papaefthymiou2014flexibility}. In this respect, ramp-up and ramp-down rates, time of availability, and start-up and shut-down response times are all components of the power provision process from conventional resources. However, the integration of VRES and other new technologies into power systems and generation plans increases uncertainty and the need for flexibility~\cite{papaefthymiou2014flexibility}.

The impact of VRES integration into power and energy systems creates a merit order effect due to lower marginal costs of production. Due to their low marginal costs of generation, the VRES plant bids enter the market merit order list before the conventional power plants bids~\cite{morales2013integrating}. Furthermore, due to the stochastic nature of VRES, their high provision of power can create supply-demand imbalance in the market. Supply-side flexibility resources, such as hydro power plants, may benefit from the imbalance by providing flexibility to restore the balance~\cite{morales2013integrating}.

\subsubsection{Demand-Side flexibility}
\label{ssec:demsid}

Information and communication technologies (ICTs) have made it easier to monitor and control consumption profiles in power systems. Real-time pricing and hourly pricing are important practices that can help to maintain the supply-demand balance. Close coordination between producers and consumers about pricing and supply-demand balance is necessary until storage technologies become cost-efficient.

Demand-side flexibility is characterized along the direction (ramp-up or ramp-down), its electrical power composition (differentiation between power and energy), its temporal characteristics defined by its starting time, duration (time of availability), and its location (spatiality)~\cite{eid2016managing}. The primary benefit of demand-side flexibility is its response to changes in market supply-demand balance and power quality problems with the support of end users. In this context, two resource technologies from the demand side are relevant: managed and responsive resources ~\cite{international2011harnessing}. Managed resources are mainly the results of direct control in power markets, such as by a DSO, a TSO, or an aggregator. Direct control strategies manage demand-side flexibility resources by load curtailing or shifting according to system needs. By contrast, the usage of responsive resources is based on the economic incentives of the market for changing end users’ consumption patterns. It is possible to use demand-side flexibility resources with optimal price signals via an efficient market design. Thus, real-time pricing, real-time metering, and economic incentives are crucial for motivating demand-side flexibility.

Industry, households, and aggregators are flexibility resources for the demand side~\cite{papaefthymiou2014flexibility}. In households, heating and cooling are important flexibility sources. Moreover, EVs are emerging as flexibility resources. They can shift their consumption in the short-term (grid-to-vehicle), while selling remaining electricity to the grid (vehicle-to-grid, V2G). Most often, demand-side technologies are applicable for local problems in short time intervals (e.g., voltage, network congestion). When congestion problem occurs, at distribution grid levels, demand-side resources are useful for congestion management. Thus, demand-side flexibility can improve the overall efficiency of the system~\cite{winterpackage2016}.

\subsubsection{Storage-side flexibility}
\label{ssec:storage}
Storage-side flexibility resources are important technologies for storing electricity and using it later~\cite{morales2013integrating}. Battery energy storage systems (BESS) can be categorized into centralized and decentralized storage units for flexibility provision~\cite{invade1}. Storage units provide power in time by collecting surplus power from VRES or other resources before the provision time~\cite{papaefthymiou2014flexibility}. Examples of different storage-side flexibility technologies are pumped hydroelectric storage technologies, compressed air energy storage, flywheels, power-to-gas plants, and batteries~\cite{divya2009battery}.  According to Divya and Østergaard (2009), BESS are the main storage flexibility resources. Some researchers regard EVs as battery storage technology due to their capacity for V2G, but in this paper we consider EVs are demand-side flexibility resources. From a power system perspective, storage flexibility from BESS can provide solutions on short-, medium-, and long-term time horizons~\cite{divya2009battery}.

\subsubsection{Grid-side flexibility}
\label{ssec:grid}
Grid infrastructure and reinforcements constitute grid-side flexibility. The definition of grid-side flexibility is the ability of a power grid to engage with demand variations, uncertainty in grid conditions, and changes in the power flow by using grid topology and system operators~\cite{li2018grid}. Transmission or distribution grid planning and operating may need grid-side flexibility to be efficient~\cite{nerc2010}.

The limitations of grid-side flexibility are often technical and are challenged by VRES and DERs~\cite{invade1}. However, the technical capabilities of grid-side flexibility may lead to reductions in the following respects:

\begin{itemize}
\item Thermal ratings: A higher number of DER and VRES connections, and growing demand can lead to violation of installed capacity (thermal ratings) in the network.
\item Voltage deviation: On-load tap changers (OLTCs) are controlled by automatic voltage control (AVC) schemes in the presence of high, low, and medium voltage situations for voltage preserve.
\item Fault level: The short circuit capacity of networks is subject to the thermal and mechanical constraints of the network. Interconnection of DERs and VRES can push the network to exceed short circuit capacity.
\item Reverse power flows: Having a reverse power flow makes balancing the low voltage side of the transformer harder and might cause congestion in both transmission and distribution systems.
\item Rapid voltage change: Instant increase in power output (ramping-up) might create rapid voltage changes and impact the grid.
\item Islanding: If a generator continues to the provision of power to an isolated grid part, consequently, the islanding occurs. Anti-islanding requirements are defined to sustain the distribution of electricity in the grid and prevent islanding.
\item Protection: There are three protection challenges for the grid. First, faults on the distribution might cause voltage deviations in the grid. Second, the aggregate generation could exceed the load on the distribution bus and the flow of power might turn in the reverse direction to  the transmission system. Third, a ground source from a generator could change the fault balance between the distribution feeder and the utility system.
\item Power quality: Integration of DERs and VRES might decrease power quality and cause voltage fluctuations, flicker, harmonics, and signaling.

\end{itemize}

With regard to local problems in power grids, grid-side flexibility is related to TSO-DSO interaction. Local network constraint management, voltage optimization, network restoration, and power flow stabilization are major applications of grid operations with flexible resources~\cite{cenelec2014}.

The grid-side flexibility is useful due to its physical capabilities to cope with changes in the power system. \cite{li2018grid} classified grid-side flexibility resources in two items: discrete grid-flexibility and continuous grid-flexibility. Discrete flexibility resources include network topology, transmission expansion planning (TEP), and line switching (LS). Dynamic flexibility resources include reactive power compensation using power electronics, phase angle, optimal power flow, FACTS (flexible alternating current transmission systems), and HVDC (high-voltage direct current).

\section{The Risk Dimension}
\label{sec:risk}

The risk dimension of flexibility provision is often neglected in characterization of flexible assets. Different risk profiles originate from the heterogeneity of technologies and end users. Also, due to the privacy concerns of participants (e.g., their data have commercial value), there is a lack of information in the market~\cite{zhao2009flexible}. The theoretical relation between risk and uncertainty is outside the scope of this paper, but we use the term risk to address the effects of uncertainty and how it affects the ability of flexibility assets to provide flexibility. At one end of the scale, we have firm flexibility provision with low probability of disruption of the service or failure to provide as promised (e.g., a portfolio of hydropower plants with reservoirs), while at the other end of the scale, we find flexibility services provided by a single windmill with a high probability of disruption or failure to deliver as promised.

To identify the risk, we first have to identify all uncertainty origins in the flexibility provision and their effect on the energy systems and markets. As long as we are able to measure or quantify the uncertainty of flexible resource, we can characterize its risk dimension. Since the beginning of flexibility research, most of the literature has highlighted the uncertainty in VRES generation plans. By contrast, risk management studies have emphasized either market price or trading risks. There are many sources of uncertainty and related risk profiles in energy systems and power markets. The following are examples of uncertainty types~\cite{kristiansen2004risk, blaesig2005methods, buygi2006network, fang2003new, kirschen2003demand, linares2002multiple}:

\begin{itemize}
\item VRES generation uncertainty
\item Demand uncertainty
\item Network availability capacity uncertainty and investments costs uncertainty
\item Fuel availability and cost uncertainty
\item Wholesale markets price uncertainty
\item Policies and regulations uncertainty
\item Participation uncertainty (in cases of a market-based approach)
\item Duration of the resource uncertainty.
\end{itemize}

These uncertainties affect the flexibility assets and services from different angles. Furthermore, the risks profiles of flexibility assets in markets have impacts on the market design and process of the flexibility usage. During the flexibility procurement and activation process, flexibility is employed to cope with these uncertainties and at the same time can potentially be affected by the same uncertainties.

The time dimension is strongly connected to the risk dimension. According to the results of a survey conducted by~\cite{helms2016timing}, a power market participant’s short-term planning contains a higher risk of inefficiency than their long-term planning. For example, many market participants conduct their trading agreements months ahead and sometimes one year ahead, and they trade the same resources to multiple markets. If they wait until the day-ahead market or intraday market, their risk could increase due to short-term uncertainties. Similarly, the shortage risk of flexibility products could originate from the contracts and obligations that the flexibility asset owner has on different time horizons. In our case, we are concerned with the uncertainty quantification of flexibility resources and the risk of shortage during provision and activation process. In a CAISO report, the shortage of ramping flexibility is described as procuring less than the requirement~\cite{caiso2016faq}. Flexible ramping product applies to both 15-minutes and real-time market designs, for upward and downward regulation. These products are designed for situations in which there is uncertainty due to demand or renewable forecast errors. The shortage of flexibility ramping products is discussed by~\cite{navid2013ramp, abdul2012enhanced, wang2017enhancing}. Insufficient flexibility ramping capacity can increase power provision prices and create market imperfections such as supply-demand imbalance.

The risk of failing to deliver flexibility can be foreseen if a robust flexibility metric exists. \cite{lannoye2012evaluation} used a flexibility metric to calculate  the time intervals of the flexibility shortage. They introduced a metric that they named insufficient ramping resource expectation (IRRE), based upon another generation adequacy metric, the loss of load expectation (LOLE). IRRE is the expected number of observations when there is a problem with the power system in the presence of forecasted or not forecasted changes in the load profile. The calculation of IRRE can represent the probability of the system coping with a shortage of flexibility. Moreover, IRRE measures individuals and the system flexibility probability.~\cite{lannoye2012power} state that there have not been any studies of the risks of the resource duration time (Figure~\ref{fig:time}). Consequently, the risk dimension needs to be addressed on an individual and resource basis according to time and spatiality dimensions.

Another type of risk associated with demand-side uncertainty is the \textit{rebound effect}~\cite{berkhout2000defining} which is also known as the payback effect~\cite{esmat2016congestion}. We can observe the rebound effect in the demand profile of a power system when the demand-side participation exits. For example, during peak hours, a demand-side participant could decrease its consumption in the grid and remove the possibility of network congestion. During off-peak hours, the same participant might increase its consumption due to lower prices to charge an EV or a battery. This behavior shows an increase in the demand profile and is subject to the possibility of congestion in the distribution grid. In this regard, the main problem is not the amount of demanded power, but the time of the demand. The uncertainty of rebound effect occurrence creates a risk to the security of supply in later periods (short-term).

System operators (DSOs, TSOs/ISOs (independent system operators)) are subject to the risk. As shown in Table~\ref{tab:serv1} and Table~\ref{tab:serv2}, the services that they provide are subject to grid congestion, shortage of flexibility, and market price risk, jointly.

An aggregator stands connected with DSOs to aggregate households’ assets in order to reduce its risk in the system or market. In a similar way to the system operator’s risk profile, the risk profile of an aggregator is a combination of  all four dimensions under discussion (i.e., time, spatiality, resource, and risk). An aggregator has many flexibility providers with different resources, spatiality, timing, and risk profiles. Therefore, an optimal portfolio of assets is important for an aggregator because the risk profiles of individuals have an impact on overall risk. To ensure its flexibility supply process, an aggregator needs to find an optimal number of assets in its portfolio based upon risk, resource technology, and spatiality and time dimensions.

\section{Flexibility products}
\label{sub:prod}

Flexibility services and products are identified by~\cite{villar2018flexibility} as the flexibility offered by a participant (e.g., an aggregator) to a market. The products offered to the TSO for system flexibility (ancillary services) usually are provided by a balance responsible party (BRP), such as CHP, hydropower plants (dispatchable), or zonal interconnections (energy products), which are defined as supply-side flexibility. The products offered to DSOs are mainly for local supply-demand balancing, voltage correction, or grid congestion management by the demand-side, storage-side, or grid-side flexibility resources (these products could also be offered by the supply-side flexibility). Furthermore, in existing market designs, power-based products such as demand-side, storage-side, and supply-side resources have shorter duration than capacity products such as grid-side flexibility.

\subsection{Product examples}
Real-life examples of ISO flexibility products are the ramping products in California ISO (CAISO). In CAISO, flexibility products, which are named \textquotedblleft flexiramp" products by~\cite{wang2016real}, should be gathered from supply-side resources in the short-term (i.e., less than minutes). In the CAISO market such products are primarily used for correcting the difference between forecasted demand and realized demand without using major energy providers~\cite{xu2012flexible}. There is no bidding for flexiramp products, due to the zero variable cost assumption of the generators that provide the products. Other markets in the USA have similar products based on ramping rate (e.g.,~\cite{navid2013ramp, parker2015ramp}), although their market settlement rules are different. Flexiramp products aim to achieve two goals: first improvement in the expected cost (market efficiency) of energy schedules; and second, the provision of incentives for generators to consider the value of ramping in both operating and investment decisions. Generators do not provide price bids for a ramping product, so prices are based just on the marginal opportunity cost of diverting capacity from energy or ancillary services to meet the ramp requirements.

Another example of a flexibility product is the DS3 plan from Ireland and its 14 products (flexible DS3) designed to meet system scarcities~\cite{flynn2016renewables}. Ireland’s TSO uses very short-term (2–10 seconds) products for frequency fixing, reactive power correction, ramping products, primary, secondary and tertiary reserves, and dynamic reactive response. Moreover,~\cite{flynn2016renewables} point out that TSO-DSO interaction is important for planning and operating of the network. Furthermore, in France, the TSO proposes capacity contracts as a quantity-based market-wide mechanism to cope with increasing peak demand and to incentivize demand-side flexibility usage for all consumers with regard to their consumption~\cite{2014french}.

In the case of DSOs, products show more variety since they include DERs. The reason for using these products is not just for market supply-demand balance but also for congestion management, voltage correction, and loss coverage~\cite{villar2018flexibility}. Principally, the flexibility is presented in the distribution grid, but it is often used in the transmission grid.~\cite{ottesen2016prosumer, ottesen2018multi, roos2014modeling} propose approaches whereby an aggregator participates with multiple flexibility resources in the distribution grid in addition to bidding in the wholesale market.

Allocation of local and system-wide resources for flexibility is important for the distribution grid and cooperation between the TSO and the DSO~\cite{smartnet2016basic}. According to~\cite{villar2018flexibility}, flexibility products are provided to local flexibility markets with DERs and other flexibility products to address grid operation issues. Many attempts to establish local flexibility markets in the industry have been reported in the literature. For example, NODES marketplace is a universal platform for local flexibility trading~\cite{nodes2018}. In distribution grids, with a pay-as-bid auction design, the NODES marketplace solves congestion problems by using continuous trading.~\cite{zhang2013flex} propose an aggregator-based local flexibility market with a flexibility clearing house (FLECH) market to promote DER for active participation in trading flexibility services. In a FLECH market, the DSO or sometimes the TSO acts as a flexibility buyer. In a FLECH design, there are three trading products: bilateral contracts, auctions, and the supermarket. In the market design, activation time, duration, and location are important for the product type. A FLECH design is aligned between the DSO and the aggregator interconnection~\cite{torbaghan2016local}. investigated the usage of prosumers’ flexibility in a decentralized perspective and found that the local market structure trades flexibility and solves problems by cost-minimizing objectives. The aim of their research is to solve distribution grid problems before using the wholesale markets.

Flexibility products can be designed as a combination of different flexibility technologies for a common purpose such as to fix voltage deviations or for congestion management.~\cite{stacking2018usef} combine flexibility from different providers for the purpose of congestion management in wholesale markets. Their product, \textit{flexibility value stacking} is based on multiple flexibility providers, who are combined either in a pool market design or a portfolio by an aggregator for trading in wholesale and balancing markets for congestion management. Flexibility value stacking products are designed as time-based, pooling/portfolio based, and double serving based.

\subsection{Flexibility product design}
\label{ssec:design}

The structure and purpose of flexibility products originates from the need for an efficient system and market design. In existing market designs, the time dimension determines the economic benefit of a flexibility product in relation to the resource dimension and technology dimension. Many existing flexibility product initiatives are system-wide products and therefore the spatiality dimension of the products is not considered~\cite{caiso2016faq, xu2012flexible}.

Flexibility service providers are heterogeneous along our four dimensions. Products may have different cost profiles for different time dimensions (activation time and duration). This leads to a need to consider the optimal alignment of markets where products can be traded. In the time dimension of flexibility (discussed in subsection~\ref{ssec:time}), the properties of the time dimension such as ramping rate and duration are relevant. When designing a flexibility product, essential qualities are how quickly fast a flexibility asset will respond to the system operator and for how long it can provide power.

In an imaginary setting, two flexibility ramping products can be considered: the first has a 5-seconds activation time and the second a 20-seconds activation time as their sweet spot in terms of cost, but both can work in a 5-second or a 20-second activation time prior to physically delivery.

\begin{figure}[!tbp]
  \centering
  \begin{minipage}[b]{0.48\textwidth}
    \includegraphics[width=\textwidth]{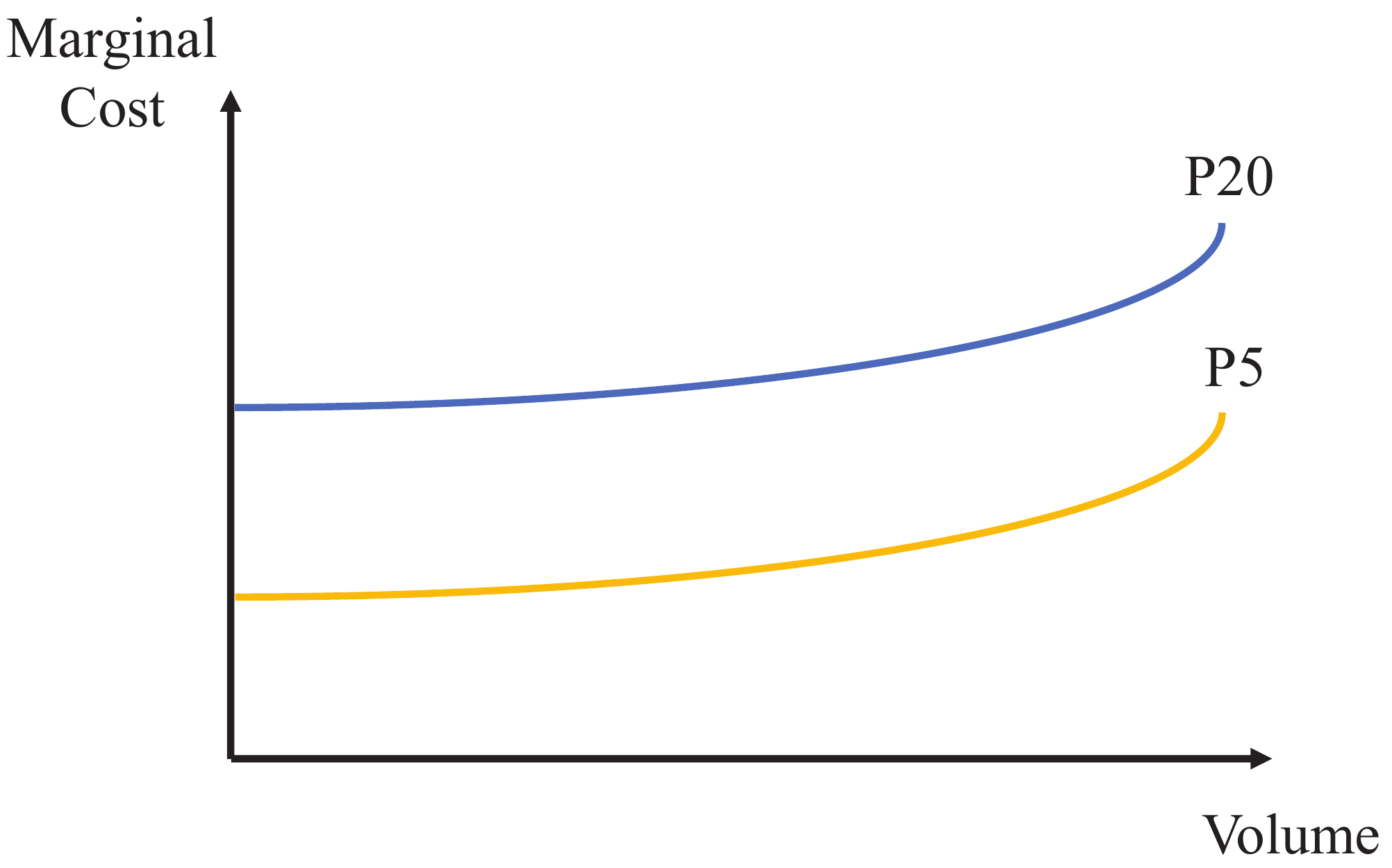}
    \caption{Cost of flexibility products in 5-second market  (graphs are in same scale).}
    	\label{fig:5sec}%
  \end{minipage}
  \hfill
  \begin{minipage}[b]{0.48\textwidth}
    \includegraphics[width=\textwidth]{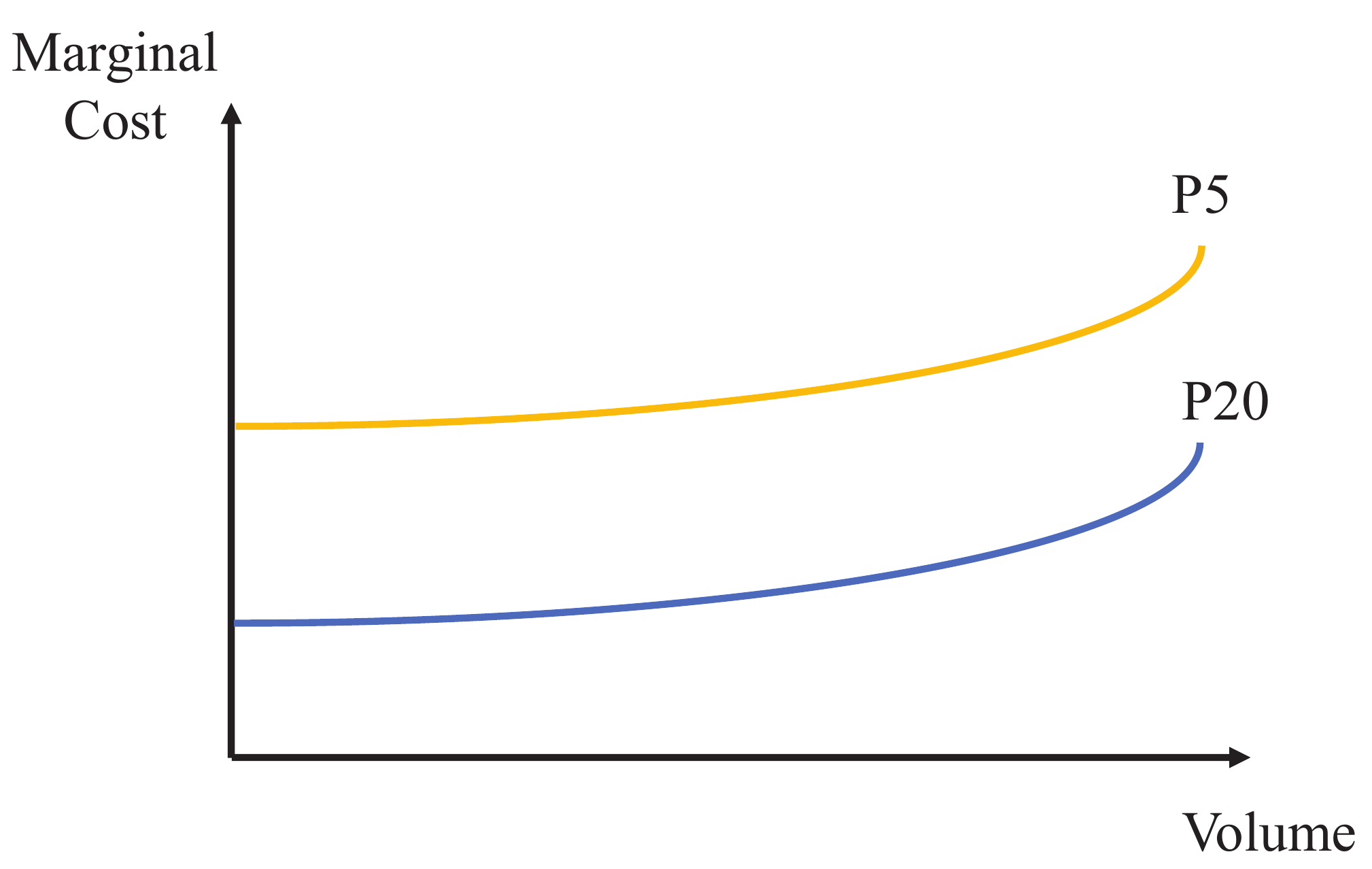}
    \caption{Cost of flexibility products in 20-second market  (graphs are in same scale).}
  	\label{fig:20sec}%
  \end{minipage}
\end{figure}

The resource with a 5-second activation time will always have lower marginal costs for the 5-second services than the 20-second resources. Similarly, the 20-second resource is better than 5-second resource for a 20-second flexibility service. If the operator dispatches 20-second technologies in 5-second markets, the operator will lose the efficiency of using flexibility. This economic viewpoint is illustrated in Figure~\ref{fig:5sec} and Figure~\ref{fig:20sec}, where P5 and P20 represent 5-second and 20-second flexibility resources, respectively. Still, it is not practical or economically efficient to prepare a market design for each asset type or resource. Therefore, the optimal market design needs to address differences in product designs for market and trading efficiency.

Similarly, one could choose optimal spatial resolution when establishing marketplaces to procure power, energy, or capacity. TSO-DSO coordination would be needed, as there would be local and non-local optimal resource allocation.

In the CAISO market, the demand curves are calculated every hour independently according to the market design (5-minute or 15-minute market) and direction (ramping-up or ramping-down). Besides the system-demand curve, there are different demand curves for each region with market imperfection~\cite{xu2012flexible}. In the case of Ireland, there are markets for an inertial response (0 to 5 seconds), reserve (5 seconds to 20 minutes), and ramping (20 minutes to 12 hours).

An aggregator or a flexibility operator chooses the flexibility resource with regard to its abilities to be dispatched in the market. For example, in CAISO, flexibility resources and technologies are dispatched or disqualified from the provision of flexible ramping products according to regulations and their technology characteristics~\cite{xu2012flexible}. Consequently, resources for shorter time intervals and longer time intervals can be separated from each other according to their technologies, either by the system operator or by the aggregator.

In addition to the time and resource dimensions, the spatiality dimension needs to be considered in flexibility product design. For example, Irish and CAISO products have system-wide initiatives in their ISOs and TSOs~\cite{caiso2016faq, xu2012flexible, irish2017}. With regard to CAISO products, the flexibility ramping products are designed as system-wide products (e.g.,~\cite{caiso2016faq}). However, the ISO might apply some regional constraints according to the problem (e.g., congestion) in the power system. In the case of Ireland, flexibility providers are spatially clustered and they generate a cost-effective strategy for grid operations (e.g.,~\cite{irish2017}).

\section{Markets for trading flexibility}
\label{ssec:markets}

To provide incentives for exploiting the value of flexibility from end users and generators, an efficient market design is essential. It is possible to provide price signals for flexibility assets in existing market designs, but they might not be sufficient. The efficiency of the existing power market designs, especially intraday (ID) and day-ahead (DA) markets for flexibility pricing could be analyzed along our four dimensions. Moreover, the spatiality dimension of flexibility refers to how to use local flexibility markets in distribution grid operations.

\subsection{Pricing flexibility in power and energy markets}
\label{ssec:trading}

The ID market design is one of the major market designs to trade flexibility and incentivize flexibility resources~\cite{pape2018impact}. In longer trading horizons (e.g., 1 week), ID market prices often are close to DA market prices. This convergence has led some researchers to disregard the importance of having separate flexibility markets~\cite{garnier2014day}. However, in their studies of flexibility pricing they have not conducted analyzes along the dimensions as those we introduce at this paper. Especially, the time dimension of flexibility resources and their spatial differences are not addressed explicitly.

In some energy-only market designs, the flexibility is withheld for peak load hours by flexibility providers~\cite{harvey2001market}. Many flexibility providers expect to recover their investment costs by trading their flexibility in peak hours. This strategy is in line with the findings of our research, such as the use of flexibility for deferring grid investments and recovering investment costs.

\cite{hoschle2017electricity} describe the pricing and market mechanism for flexibility trading in the presence of price caps and cost recovery conditions. Price caps in energy markets lead to higher prices; hence, trading flexibility in peak hours increases the power prices (ramp-up in scarcity hours). Price capping might be an option for market mechanisms, but price cap revenues are related to revenues from flexibility trading. The major cost recovery for flexibility investments comes from earnings from trading pricing at peak pricing periods instead of off-peak or regular trading periods. Naturally, prices for flexibility are mainly affected by (marginal) costs of technologies and the applied price cap.

From the spatiality dimension perspective, aggregators can access different resources in various locations. Hence, integration of different market zones, both in time and spatiality, could increase the flexibility in the system if the products traded are relevant over a large geographical area (e.g., active power in non-congested grids). The combination of different areas and generators in the same market leads to better allocation of reserves and reduces the costs of marginal generation, especially for supply-side flexibility resources~\cite{nicolosi2010wind, riesz2015designing, grande2008exchange, farahmand2012balancing}.

Another problem with existing market pricing mechanisms is the lack of incentives for the flexibility providers of flexibility activation, in times of both power scarcity and power surplus. In situations when flexibility is provided from demand-side resources, we can observe that market power shifts from the generators to the end users of electricity~\cite{lund2015review}.~\cite{su2009quantifying} show that demand-side flexibility resources and their bids can outperform conventional price bids and reduce flexibility prices. Therefore, it is essential to incentivize the demand-side flexibility in a market design.

Flexibility pricing examples from CAISO, MISO (Midcontinent Independent System Operator), and SPP (Southwest Power Pool)  markets indicate that the flexibility products are subject to DA market and real-time (RT) pricing. In these markets, flexibility is characterized by considering mainly the time dimension (i.e., ramping rate)~\cite{navid2013ramp, wang2016real, xu2012flexible, parker2015ramp}. Another example is the Irish TSO, which proposes products by considering the spatiality, resource, and time dimensions, but mainly emphasizes the time dimension because of system needs. In EirGrid, the pricing of fourteen different products is ideally done under real-time pricing~\cite{flynn2016renewables, irish2017}. In the French TSO case, capacity obligations and certificates construct the price mechanism, especially for peak-hours electricity provision~\cite{2014french}.

\subsection{Local flexibility markets}
\label{ssec:local}
Power markets are fundamentally different because the needs of participants are numerous with regard to resource technology, time of availability, geography, and risk preferences. The introduction of the entity “prosumer” to the power and energy markets changes power market designs. The change in market designs from centralized to decentralized, and the integration of prosumers into existing markets is investigated by~\cite{parag2016electricity}, with respect to four structural attributes: the peer-to-peer model, prosumer-to-islanded microgrids, prosumer-to-interconnected microgrids, and the organized prosumer group model. In the peer-to-peer model, prosumers are directly interconnected with each other for buying and selling power and energy from others. In prosumer-to-interconnected microgrids, prosumers provide their services to a microgrid that is a part of a larger grid. The prosumer-to-islanded microgrids comprise prosumers who provide services to independent, non-interconnected microgrids. In the fourth and final market structure, organized prosumers create a pool among themselves and trade with each other.

Each market design typology, whether pooled or bilateral, has different attributes along our four dimensions. A local and consumer-centric market design, such as a local flexibility market (LFM), might be an efficient market design for the flexibility pricing and trading.

In order to design a local flexibility market, a general list of market design principles needs to be followed before introducing the details of the flexibility trading. According to~\cite{newbery2018market}, six principles of a good market design are as follows:

\begin{enumerate}
\item Correct the market as quickly as possible in cases of failure. By reducing the reliability on subsidiarity, the market imperfection will be corrected as soon as possible.
\item Allow for appropriate cross-country variation in market design. Ensuring the security of supply is a local issue.
\item Use price signals and network tariffs to represent the value of electricity provision services. Include the provision of flexibility. This principle has long and short-term effects such as deferring the investments and sustaining the efficient dispatch.
\item Collect network fixed costs from the market. The difference between efficient prices and regulated prices allows for revenue from end users.
\item Provide incentives for low carbon investment. Provide efficient risk-averse financing for low-carbon and capital-intensive investments in electricity markets.
\item Retain the flexibility to respond to changing information in the market, such as information relating to lower costs and different technologies.
\end{enumerate}

In addition to the six fundamental principles of a market design, a local flexibility market requires incentives for the valuing flexibility. For stronger incentives to exploit flexibility from end users and to increase efficiency in the market and systems, LFMs are crucial on specific grid or market purposes. The need for an LFM is specific to each case. The majority of researchers consider the need for LFMs as decentralized and separate from wholesale markets. In some cases (e.g.,~\cite{heinrich2020ecogrid}), they suggest that an LFM should complement the balancing markets. According to Jin, Wu, and Jia (2020), recent studies have provided good insights into an efficient market design for flexibility trading~\cite{jin2020local}. In addition, LFMs are useful for various services, such as market-oriented services, system-oriented services, and grid-oriented services~\cite{minniti2018local}.

To design an LFM for pricing and trading flexibility, the market design needs to address our four dimensions. According to the flexibility service, for example the voltage deviation service, the market considers the spatiality of the flexibility resource because the voltage needs to be fixed at certain locations in the grid topology (active and reactive power distribution). With regard to another flexibility service, namely congestion management, it is important to address the congestion with the correct timing (peak load time); this refers to the time dimension of the flexibility. In case of risk dimension, the LFM is requires to cope with the market liquidity risk in order to provide sufficient amount of power from flexible resources (scarcity of flexibility). Hence, the LFM design needs to be shaped with respect to the risk dimension discussed in this paper. However, the TSO-DSO coordination and the coexistence of different LFMs have to be considered for higher efficiency for flexibility usage.

\subsection{The need for TSO-DSO services and coordination based on flexibility}
\label{sec:dsotso}

A system-wide approach to coordination among multiple market participants and operators is needed for reliability and efficiency of the power system. DSOs can deal with local problems by flexibility trading, while TSOs manage TSO-DSO interaction~\cite{villar2018flexibility, birk2016tso, entsoe2015dso, hansen2013coordination, zegers2014tso}. Accordingly,~\cite{smartnet2016basic}, suggest five different coordination models: centralized ancillary services market, local ancillary service market, shared balancing responsibility, common DSO-TSO ancillary service market, and integrated flexibility market. According to the~\cite{smartnet2016basic}, in the centralized ancillary services model, a single market with only a TSO as buyer is designed without the participation of the DSO. In the local ancillary service market model, the DSO is the user of the local flexibility and establishes a local market. The shared balancing responsibility model indicates that the local markets have to provide lower entry barriers to DERs for TSO-DSO coordination. In a common TSO-DSO ancillary services market model, the TSO and the DSO collaborate to use flexible resources optimally. Lastly, the integrated flexibility model both increases the possibilities for BRPs to solve supply-demand imbalances, and increases the market liquidity.

The provision of flexibility services by the TSO and DSO are related to voltage, congestion, balancing, black-start, and interoperability for coordinated protection~\cite{zegers2014tso}. There are ongoing discussions about pricing these services based on flexibility assets, as we have mentioned in subsection~\ref{ssec:trading}. System services that are provided by the DSO and the TSO (or ISO) are listed in Table~\ref{tab:serv1} and Table~\ref{tab:serv2}, according to~\cite{birk2016tso} and~\cite{hansen2013coordination}.

	\begin{table}[ht!]
\centering
		\caption{TSO and ISO services and pricing mechanisms~\cite{birk2016tso, hansen2013coordination}.}
		\resizebox{\linewidth}{!} {
			\begin{tabular}{lll}
				\hline
				\textbf{TSO/ISO services}          & \textbf{ISO pricing}                                      & \textbf{TSO pricing}                                      \\ \hline
				Electrical energy                  & \multirow{3}{*}{Local marginal prices (LMPs)}                                     & \multirow{2}{*}{Zonal}                                    \\ \cline{1-1}
				Transmission energy losses         &                                                           &                                                           \\ \cline{1-1} \cline{3-3} 
				Transmission congestion            &                                                           & Congestion management markets                             \\ \hline
				Reserves                           & Co-optimized with LMPs                                    & Balancing markets                                         \\ \hline
				Reactive power and voltage control & \multirow{2}{*}{Regulated prices and bilateral contracts} & \multirow{2}{*}{Regulated prices and bilateral contracts} \\ \cline{1-1}
				Black-start                        &                                                           &                                                           \\ \hline
			\end{tabular}
		}
		\label{tab:serv1}
	\end{table}

	\begin{table}[ht!]
\centering
		\caption{DSO services and pricing mechanisms~\cite{birk2016tso, hansen2013coordination}.}
		\resizebox{\linewidth}{!} {
			\begin{tabular}{ll}
				\hline
				\textbf{DSO services}              & \textbf{Pricing}                                                \\ \hline
				Electrical energy                 & \multirow{4}{*}{Regulated or competitive retail supply tariffs} \\ \cline{1-1}
				Distribution energy losses         &                                                                 \\ \cline{1-1}
				Distribution congestion           &                                                                 \\ \cline{1-1}
				Reactive power and local voltage control
				&  
				\\ 
				\cline{1-1} 
				Peak shaving &  \\ \hline
				Network connection and reliability & \multirow{2}{*}{Averaged network tariffs}                       \\ \cline{1-1}
				Network deferral                   &                                                                 \\ \hline
			\end{tabular}
		}
		\label{tab:serv2}
	\end{table}

\subsubsection{Interaction along flexibility resource}
\label{ssec:intres}

The distinction between DSO and TSO services in Table~\ref{tab:serv1} and Table~\ref{tab:serv2} originates from the voltage and frequency requirements of the system. The TSO considers frequency and grid congestion issues whereas the DSO focuses on voltage deviation, grid congestion, and losses issues. The requirements of frequency deviations for conventional resources (supply side) is much stricter than requirements for demand-side resources. The reason for this is that the voltage should be higher when electricity is injected into the grid from the supply side but should be lowered when it reaches end users for utilization  (high-voltage to low-voltage grid). Therefore, local resources managed by the DSO have different voltage requirements compare with the non-local resources owned by the TSO. As a result, besides voltage and frequency challenges, the congestion management for an entire grid is diversified by the DSO and TSO concerning their local flexibility and grid resources. TSO and DSO services can differ because their products (e.g., flexibility resources) can differ.

According to~\cite{birk2016tso}, DSO and TSO services can compete with each other within the same level of the grid. Moreover, flexible power resources can compete in DA, ID or balancing markets as either energy or power, but not as capacity. Flexibility resources should be bid to markets that are most profitable for them. Furthermore, for flexibility trading, the bidding process should provide optimal incentives and price signals for market participants to continue~\cite{birk2016tso}. In this regard, the reduction of market barriers would be helpful, as stated by~\cite{macdonald2012demand}.

\subsubsection{Interaction along spatiality}
\label{ssec:intspat}

To coordinate flexibility resources, system operators should communicate with each other according to their spatial responsibilities. The spatial differences among flexibility assets have impacts on their technology and their mitigation of grid problems~\cite{birk2016tso}. Resources that are located in different geographies, as illustrated in Figure~\ref{fig:spatial}, have different incentives, technologies, contracts, and market power. In particular, we cannot expect flexibility resource from a transmission level (high-voltage) to act in a similar way to a small demand-side resource in a distribution grid.

The congestion management service is common in both types of system operators (i.e., TSOs and DSOs) and is increasing in importance due increases in local power generation. DSOs can use demand-side and storage-side flexibility resources for local congestion management, whereas TSOs can use supply-side and grid-side flexibility for transmission grid services. These facts stress the coordination of flexible resources. The geographical information tags for DSO and TSO market bids are presented by~\cite{entsoe2015dso} for the coordination of flexible resources.

\subsubsection{Interaction along time}
\label{ssec:dimtime}
Flexibility assets can provide long-term and short-term solutions for markets and services. Furthermore, a short-term resource can bid for a long-term perspective, and at any point in time there might be conflict (or overlap) among the contracts. A DSO could use its resources for local voltage balancing, while a TSO might want the same resources for congestion management in the grid. Such situations need a high level of coordination between the TSO and DSO. As shown in Table~\ref{tab:serv1} and~\ref{tab:serv2}, the DSO and TSO provide different services, but both provide services for grid congestion management.

The coordination of the DSO and TSO should be evaluated in two time periods, such as short term and long term. Currently, there is an ongoing TSO-DSO coordination in long-term planning  in the literature and in the industry. Smart grid initiatives, network expansion planning, and research programs are examples of long-term collaboration~\cite{birk2016tso, entsoe2015}. However, the coordination between the DSO and TSO should include short-term solutions for congestion, voltage, and frequency problems in further consideration of new market designs.

\subsection{Need for change in existing power markets}
\label{sec:evol}
The integration of VRES and the transition of energy systems affect the management, technology, and economics of market designs and power systems from centralized to decentralized, and from a regulated structure to a deregulated structure~\cite{eid2016managing}. From a centralized to decentralized perspective, the market scale is downsized from a national design to local market design. The resources available in national markets are still valid for use at the local scale, but it would be problematic to use certain flexibility technologies due to their market power, amount of power produced, and time of availability. Therefore, for flexibility trading, there is a need for change in market designs from national to local scale.

A comparison of existing market designs and their participant profiles is important in order to understand the need for change in market designs. Using DERs and VRES increases the risk for power markets and systems due to the uncertainty in generation and consumption profiles. Depending on the market design, the risk can be reduced. A change in power markets needs to include the risk profiles of intermittent resources in order to increase efficiency.

A structural comparison of flexibility provision in the current market situation and a basic understanding of the need for change in power systems and markets is presented in Table~\ref{tab:compare}. Originally,~\cite{eid2016managing} studied a similar version of this table with only DERs. For this reason, we propose an extension with all flexibility technologies, in a time-coupled context, considering our four dimensions, in addition to flexibility products and related market mechanisms. Our novel expansion is the introduction  of uncertainty and risk in Table~\ref{tab:compare}. 

All dimensions discussed in this paper are incorporated in Table~\ref{tab:compare}. In addition to the four dimensions, the table~\ref{tab:compare} shows existing market mechanisms, flexibility products, and the connection type of the grid. Time interval presents the availability of flexibility resources according to the time dimension. Some of these resources exist in multiple markets. The type of product is mainly related to the flexibility resource technology. The technology of the resource is related to its location and connection to the grid. Distribution grid (DSO) technologies are used for local purposes, whereas transmission grid (TSO) technologies are used for non-local and local reasons.

\begin{sidewaystable}[p]
	\centering
	\resizebox{\textwidth}{!}{
		\begin{tabular}{@{}llllllllll@{}l}
			\hline
			\multicolumn{5}{l}{\textbf{Time interval}}                                                                                                                                                                                           & \textbf{\begin{tabular}[c]{@{}l@{}}Market \\ mechanism\end{tabular}}        & \textbf{\begin{tabular}[c]{@{}l@{}}Product \end{tabular}} & \textbf{\begin{tabular}[c]{@{}l@{}}Flexibility \\ provider\end{tabular}}                                  & \textbf{Spatiality}                                                             & \textbf{\begin{tabular}[c]{@{}l@{}}Connection \\ to \\ grid\end{tabular}}       & \textbf{Uncertainty}                                                                                                                                                              \\  \midrule
			\multicolumn{1}{l}{Real-time} & \multicolumn{1}{l}{\multirow{8}{*}{Within day}} & \multicolumn{1}{l}{\multirow{19}{*}{Short term}} & \multicolumn{1}{l}{\multirow{33}{*}{Medium term}} & \multicolumn{1}{l}{\multirow{45}{*}{Long term}} & \begin{tabular}[c]{@{}l@{}}Direct \\ control\end{tabular}                    & \multirow{2}{*}{Power}                                                           & Household appliances                                                                                      & Local                                                                           & Distribution                                                  & \multirow{2}{*}{\begin{tabular}[c]{@{}l@{}}Resource duration, \\ Demand,\\ Congestion\end{tabular}}                            \\ \cmidrule(r){1-1} \cmidrule(lr){6-6} \cmidrule(l){8-10} 
			\multicolumn{1}{l|}{}           & \multicolumn{1}{l}{}                        & \multicolumn{1}{l}{}                            & \multicolumn{1}{l}{}                          & \multicolumn{1}{l}{}                           & \begin{tabular}[c]{@{}l@{}}Indirect \\ control\end{tabular}                  &                                                                                     & \begin{tabular}[c]{@{}l@{}}Household appliances, \\ EVs\end{tabular}                                      & Local                                                                           & Distribution                                                                           \\ \cmidrule(r){1-1} \cmidrule(l){6-11} 
			\multicolumn{1}{l|}{}           & \multicolumn{1}{l}{}                        & \multicolumn{1}{l}{}                            & \multicolumn{1}{l}{}                          & \multicolumn{1}{l}{}                           & \begin{tabular}[c]{@{}l@{}}Balancing \\ markets\end{tabular}                 & \multirow{2}{*}{\begin{tabular}[c]{@{}l@{}}Energy and \\ power\end{tabular}}     & \begin{tabular}[c]{@{}l@{}}EVs, Industrial DS, \\ Aggregators\end{tabular}                                & \multirow{7}{*}{\begin{tabular}[c]{@{}l@{}}Local and \\ non-local\end{tabular}} & \multirow{7}{*}{\begin{tabular}[c]{@{}l@{}}Transmission, \\ Distribution\end{tabular}} & \multirow{4}{*}{\begin{tabular}[c]{@{}l@{}}Resource duration, \\ Demand,\\ Congestion, \\ Fuel availability and cost, \\ Wholesale market price, \\ VRES generation\end{tabular}} \\ \cline{1-2} \cline{6-6} \cline{8-8}
			& \multicolumn{1}{l|}{}                        & \multicolumn{1}{l}{}                            & \multicolumn{1}{l}{}                          & \multicolumn{1}{l}{}                           & \begin{tabular}[c]{@{}l@{}}Ancillary \\ services\end{tabular}                &                                                                                     & \begin{tabular}[c]{@{}l@{}}Aggregators, Conventional, \\ Renewable\end{tabular}                           &                                                                                 &                                                                                        \\ \cmidrule(r){1-2} \cmidrule(lr){6-8}
			& \multicolumn{1}{l|}{}                        & \multicolumn{1}{l}{}                            & \multicolumn{1}{l}{}                          & \multicolumn{1}{l}{}                           & Intraday                                                                     & Energy                                                                              & \begin{tabular}[c]{@{}l@{}}Aggregators, Conventional, \\ Renewable\end{tabular}                           &                                                                                 &                                                                                        \\ \cmidrule(r){1-2} \cmidrule(lr){6-8}
			& \multicolumn{1}{l|}{}                        & \multicolumn{1}{l}{}                            & \multicolumn{1}{l}{}                          & \multicolumn{1}{l}{}                           & Day-ahead                                                                    & Energy                                                                              & \begin{tabular}[c]{@{}l@{}}Aggregators, Conventional, \\ Renewable, Storage\end{tabular}                  &                                                                                 &                                                                                        \\ \cmidrule(r){1-3} \cmidrule(lr){6-8}\cmidrule(lr){11-11}
			&                                              & \multicolumn{1}{l|}{}                            & \multicolumn{1}{l}{}                          & \multicolumn{1}{l}{}                           & \begin{tabular}[c]{@{}l@{}}Forward \\ markets\end{tabular}                   & \begin{tabular}[c]{@{}l@{}}Energy and \\ power\end{tabular}                      & \begin{tabular}[c]{@{}l@{}}Conventional, Renewable, \\ Storage\end{tabular}                               &                                                                                 &                    & \begin{tabular}[c]{@{}l@{}}Demand, \\ Fuel availability and cost, \\ Wholesale market price, \\ VRES generation\end{tabular}                                                                    \\ \cmidrule(r){1-3} \cmidrule(lr){6-8}\cmidrule(lr){11-11}
			&                                              & \multicolumn{1}{l|}{}                            & \multicolumn{1}{l}{}                          & \multicolumn{1}{l}{}                           & \begin{tabular}[c]{@{}l@{}}Capacity \\ markets\end{tabular}                  & Capacity                                                                            & Conventional, Renewable                                                                                   &                                                                                 &                                                    & \begin{tabular}[c]{@{}l@{}}Demand, \\ Fuel availability and cost, \\ Wholesale market price\end{tabular}                                     \\ \cmidrule(r){1-4} \cmidrule(lr){6-8}\cmidrule(lr){11-11}
			&                                              &                                                  & \multicolumn{1}{l|}{}                          & \multicolumn{1}{l}{}                           & \begin{tabular}[c]{@{}l@{}}Network expansion \\ and investments\end{tabular} & Capacity                                                                            & \begin{tabular}[c]{@{}l@{}}Network reconfiguration, \\ Grid expansion, \\ Capacity expansion\end{tabular} &                                                                                 & & \begin{tabular}[c]{@{}l@{}}Network investments, \\ Policy and regulation\end{tabular}   \\ \bottomrule
		\end{tabular}
	}
	\caption{Structure of flexibility trading in current systems and market designs.}
	\label{tab:compare}
\end{sidewaystable}

For some flexibility technologies, it is possible to use the dimensions in multiple market designs and time scales as shown in Table~\ref{tab:compare}. However, the risk profiles and spatiality of flexibility resources indicate a market design that is dimension-specific, especially for small-scale technologies such as demand-side and storage-side flexibility resources. In addition, throughout different market designs, the risk has different impacts on market participants and their choice of market for trading flexibility. Hence, there could be changes in existing market designs to include more flexibility resources, in addition to designing local (flexibility) markets by considering the discussed four dimensions.

\section{Conclusion and Outlook}
\label{sec:conc}
There are still many topics to be discussed in flexibility in power and energy markets. Various market designs in different countries need different models and solutions. Unfortunately, a unique model does not exist. All energy systems have their own characteristics, and therefore we see different spatiality, resources, time intervals, and risk profiles.

In this paper, we have characterized flexibility along four dimensions. We have introduced flexibility resources, their possible geographical locations, time constraints, and risk profiles during procurement and trading process. Connections between these dimensions have been discussed theoretically. We have presented products and services from different countries  with flexibility trading systems, focusing on products, pricing, and valuation markets, and how they support the flexibility products and services that are needed to balance the power markets supply and demand side on different horizons and solve other system problems with voltage, frequency and congestion. We have discussed pricing of the flexibility in possible local flexibility markets in addition to existing power and energy markets.

Our main contribution is the discussion of the flexibility along four dimensions and an examination of the relations between them. We have presented services and products along flexibility dimensions with different pricing methods. We argue that the flexibility services have to be procured and deployed by considering four dimensions. In conclusion, it is possible to evaluate flexibility in different market designs, but for an efficient valuation of the flexibility, a local flexibility market might be needed. Considering risk profiles and uncertainty of flexibility assets in flexibility provision could help to decrease inefficiencies in flexibility usage and local market design. At different levels in the networks, TSO-DSO coordination is essential to provide services based on flexibility.

\section*{Acknowledgments}
This work is funded by CINELDI - Centre for intelligent electricity distribution, an 8 year Research Centre under the FME-scheme (Centre for Environment-friendly Energy Research, 257626/E20). The authors gratefully acknowledge the financial support from the Research Council of Norway and the CINELDI partners.

\bibliographystyle{elsarticle-num}

\end{document}